\documentclass[aps,pra,twocolumn,superscriptaddress,10pt,longbibliography]{revtex4-2} 
\usepackage{graphicx}
\usepackage{dcolumn}
\usepackage{bm}

\usepackage[colorinlistoftodos]{todonotes} 

\usepackage[normalem]{ulem}
\usepackage[utf8]{inputenc}
\usepackage{amssymb}   
\usepackage{amsmath}   
\usepackage{braket}    
\usepackage{subfig}    
\usepackage{comment}   
\usepackage{float}	   
\usepackage{multirow}
\usepackage{color}
\usepackage{fancyhdr}
\usepackage{amsthm}
\usepackage{tabularx}
\usepackage{setspace}
\usepackage[export]{adjustbox}
\usepackage{stfloats}
\usepackage{tabu}
\usepackage{xcolor}

\newcommand{\be}[1]{\begin{equation}\label{#1}}
\newcommand{\ee}{\end{equation}}
\newcommand{\bea}[1]{\begin{eqnarray}\label{#1}}
\newcommand{\eea}{\end{eqnarray}}

\newcommand{\Fig}[1]{Fig.(\ref{#1})}

\usepackage{fancyhdr}
\begin{document}
\fancyhead[R]{\ifnum\value{page}<2\relax\else\thepage\fi}

\preprint{APS/123-QED}

\title{Microring resonator-based photonic circuit for faithfully heralding NOON states}
\author{Ryan Scott}
\affiliation{Department of Physics, Virginia Tech, Blacksburg, Virginia 24061, USA}
\author{Peter L. Kaulfuss}\email{corresponding author: kaulfuss\_peter@bah.com}
\affiliation{Booz Allen Hamilton, 8283 Greensboro Drive, McLean, VA 22102, USA}
\author{A. Matthew Smith}
\affiliation{Air Force Research Laboratory, Information Directorate, 525 Brooks Rd, Rome, NY, 13411, USA}
\author{Paul M. Alsing}
\affiliation{University at Albany-SUNY, Albany, NY 12222, USA}
\author{Wren Sanders}
\affiliation{School of Physics and Astronomy, Rochester Institute of Technology, 85 Lomb Memorial Drive, Rochester, New York 14623, USA}
\author{Gregory A. Howland}
\affiliation{School of Physics and Astronomy, Rochester Institute of Technology, 85 Lomb Memorial Drive, Rochester, New York 14623, USA}
\author{Edwin E. Hach, III}
\affiliation{School of Physics and Astronomy, Rochester Institute of Technology, 85 Lomb Memorial Drive, Rochester, New York 14623, USA}


\date{\today}

\begin{abstract}
We have designed a Micro-Ring Resonator (MRR) based device that allows for the post-selection of high order NOON states via heralding. NOON states higher than $N=2$ cannot be generated deterministically. By tuning the coupling parameters of the device we can minimize the amplitudes of the `accidental' states to maximize the probability of obtaining the NOON state upon a successful heralding event. Our device can produce a 3-photon NOON state output with 100\% certainty upon a successful heralding detection, which occurs with probability $\frac{8}{27}$ for optimal tunable device parameters. A successful heralding event allows for non-destructive time of flight tracking of the NOON state thus establishing a significantly enhanced level of engineering control for integration of the NOON state into scalable systems for quantum sensing and metrology. We further discuss extensions of our technique to even higher NOON states having $N=4,5$.
\end{abstract}

\maketitle
\thispagestyle{fancy}


Several years ago, Dowling introduced the concept of the NOON state and its potential for radical enhancement of interferometric techniques, specifically allowing for Heisenberg limited phase sensitivities, far surpassing the Standard Quantum Limit on such measurements \cite{LeeKokDowling:2002}. Stated another way, the NOON state allows for a distinct quantum advantage in the area of metrology. Shortly after the introduction of the NOON state, Gerry et. al. proposed optical parity as a specific physical observable capable in principle of optimizing the NOON state quantum metrological advantage \cite{Gerry:2003}. The $N=2$ realization of the NOON state is the output resulting from the linear interaction of identical photons with a 50/50 beam splitter; this is the well-known Hong-Ou-Mandel Effect, ubiquitous in quantum information processing, sensing, and metrological applications \cite{HOM:1987,Fearn:1987,Rarity:1989,Abram:1986,Shih:1988}. Over the years, the production and engineering control of so-called High-NOON $(N>2)$ states has proved to be a formidable challenge, largely related to the absence of nonlinear materials sufficient to mediate the interactions that would be required to generate High-NOON states deterministically. A promising strategy for emulating these otherwise elusive nonlinear interactions is provided by dilating physical systems so as to include ancillary modes upon which projective measurements are performed in ways that leave the target modes in the desired high-NOON state. The benefit from from this approach is that the local isometry induced upon the target modes via the measurement induced nonlinearity can be engineered to produce the desired output. The cost of the approach is that it is inherently probabilistic and, worse, often involves the demolition of the state one hopes to generate. This leads to a situation in which one may know that one had generated a high-NOON state prior to some post-processing, but in gaining that information one has rendered the state itself useless or even nonexistent. Post processing is a violent business.

Shortly after its introduction, Kok et al. \cite{Kok:2002} presented a basic protocol for generating High-NOON states using linear optics and conditional single-photon detections. Pryde et. al. extended this analysis by proposing a more specific, fiber-based architecture and by considering the operation of the circuit using more easily generated coherent states as opposed to twin Fock states as the input states \cite{Pryde:2003}. Even earlier, Zou et al. developed a basic scheme based upon conditional zero-photon measurements \cite{Zou:2001}. Around the same time, Fiurasek \cite{Fiurasek:2002} proposed a similar method and presented encouraging results for probabilities of success for generating the High-NOON state. All of these early proposals for generating High-NOON states rely heavily upon the conditional detection of zero photons in ancillary modes. In some cases these null detections occur externally to the individual `building blocks' of the circuit (Fiurasek, Pryde, Kok) and in others they serve more directly in the successful operation of each device in the circuit (Zou). In each of these situations, the generation of the High-NOON state is heralded by null measurements on some array of ancillary detectors. Afek et al \cite{Afek:2009} proposed an interesting scheme based upon the interference at a beam splitter between an ordinary coherent state and a squeezed state generated by Spontaneous Parametric Down Conversion (SPDC). Hofmann and Ono have shown that the beam splitter and complex amplitudes of the input state can be parametrically adjusted to deterministically produce an output state that is a linear superposition of NOON states such that the probability amplitudes are significant for several ``high" values of N \cite{Hofmann:2007}. Afek et al. demonstrated this experimentally via a system of coincidence counts and post processing that allowed them to make an ex post facto assay of the relative intensity of each High-NOON component of the output state vector form the beam splitter. The result is an important advancement in understanding the generation of highly path-entangled photonic states, but unfortunately, this approach costs the state vector itself. More recently, Zhang and Chang \cite{Zhang2018NOON} have proposed several experimental approaches for generating multi-mode High-NOON states. Their ideas, while promising, feature several steps involving SPDC and in one instance, a large cross-Kerr interaction that is unlikely to be available naturally or synthetically, especially in a deployable sensing system.   

Advances in Quantum Photonic Integrated Circuits (QPICs) over the past couple of decades have allowed for on-chip realizations of quantum optical experiments that previously required significant tracts of real estate on large optical tables \cite{Adcock:2020,Polino:2020,Lee:2024}. Presently, a chip fabricated with 50 micron-scale linear optical elements can be configured for several different experiments, or, perhaps more practically, with several sets of channels all devoted to the same photonic interaction. This latter arrangement allows for an unprecedented level of quantum photonic engineering redundancy, all within a manageable volume that is readily suited for packaging. This redundancy, in turn, paves the way for the practical, systems level deployment of linear optical circuits designed to perform probabilistic, isometric transformations on multi-mode systems. As a result, accommodation of the ancillary modes required for implementing measurement induced nonlinearities (in lieu of deterministic ones, which have proved elusive) for applications such as two-qubit gates, quantum control, and quantum state engineering have become a practical possibility \cite{Labonte:2024,Matthews:2011,Yu:2022}.

In this paper, we investigate a relatively simple, inherently scalable double-bus Micro-Ring Resonator (db-MRR) based silicon nanophotonic circuit that easily accommodates ancillary modes sufficient to support a broad range of measurement induced nonlinearity. Further, this architecture can be parametrically tuned dynamically to allow for the in situ optimization of the probability for a desired quantum photonic output. Our proposal for generating High-NOON states offers the added advantages of faithful heralding via non-zero photon detection in the ancillary modes: providing a time-of-flight time stamp on the desired NOON state and lacking the need for any post processing, thereby preserving the target state itself.

{\flushleft{\it Scheme:}}
\begin{figure}[H]
	\centering
	\includegraphics[width=3in,keepaspectratio]{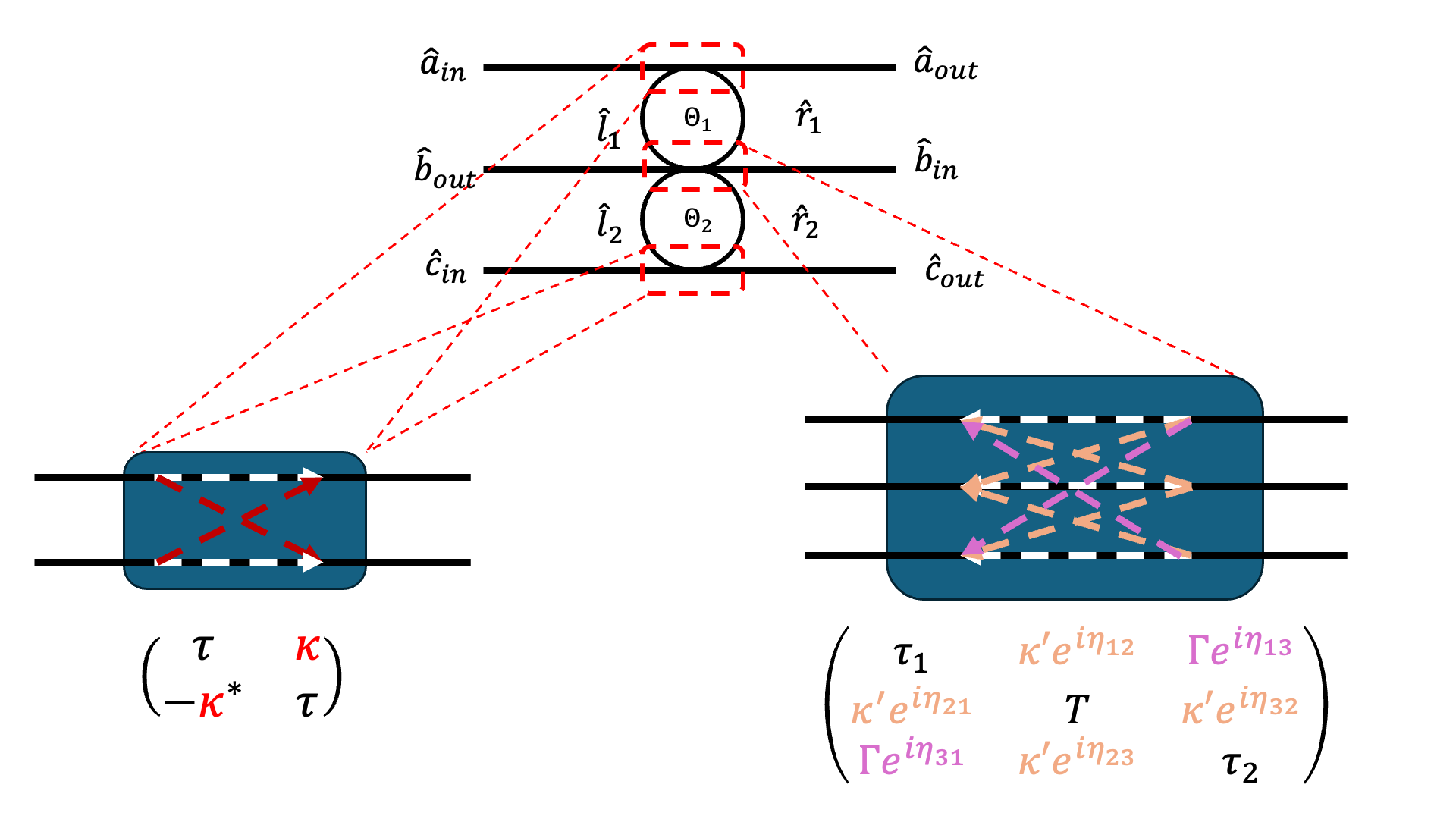}
	\caption{Schematic representation of our MRR-based NOON-state creation circuit element.}
    \label{fig:rr_circuit}
\end{figure}

We begin with the basic circuit shown in Fig. \ref{fig:rr_circuit}. The circuit consists of three parallel single-mode waveguides labeled a, b, and c coupled to two micro-ring resonators with internal phases $\theta1$ and $\theta2$, respectively. From a theoretical point of view, this will consist of a collection of three distinct systems of equations which all must be satisfied simultaneously. These can be solved in a customary manner to extract a $3 \times 3$ linear operator relating the input and output modes \cite{Hach:2014, Kaulfuss:2023Identical}, which we refer to generally as the $S$-matrix for the linear optical device.

\begin{equation}
    S=\begin{pmatrix}
    S_{11} & S_{12} & S_{13} \\
    S_{21} & S_{22} & S_{23} \\
    S_{31} & S_{32} & S_{33} 
    \end{pmatrix}
    \label{Smatrix}
\end{equation}

\noindent Here the elements of $S$ are algebraic functions of the tunable parameters characterizing the junctions. 

To derive the forms of these algebraic functions, we follow the work of \cite{Skaar:2004} and make natural assumptions about the device in Fig. \ref{fig:rr_circuit}. First, we assume that the device is symmetric in its characterizing parameters about the middle mode, meaning that the outer junctions satisfy the usual relations \cite{Hach:2014}, where the transfer matrix at each outer junction is a simple directional coupler, $U_{2\times2}$:

\begin{equation}
    U_{2 \times 2} = \begin{pmatrix}
        \tau & \kappa\\
        -\kappa^{*} & \tau
    \end{pmatrix}.
    \label{DCmatrix}
\end{equation}

In order to build the $S$ matrix we also need to know the form of the coefficients characterizing the middle mode junction. We again argue that the transitions should be symmetric (the transitions from the upper ring to the middle waveguide are symmetric with the transitions from the lower ring to the middle waveguide). Transitions directly from the upper ring to the lower ring will also be symmetric but distinct from the transitions from waveguide to ring. Finally we assume that all transmission coefficients are real and that the off-diagonal phases in the matrix are arbitrary. Under these natural assumptions, we have a local $3 \times 3$ transfer matrix specific to our system of the form, $U_{3 \times 3}$:

\begin{equation}
    U_{3 \times 3} = \begin{pmatrix}
        \tau_1 & \kappa' e^{i\eta_{12}} & \Gamma e^{i \eta_{13}}\\
        \kappa' e^{i \eta_{21}} & $T$ & \kappa' e^{i\eta_{23}}\\
        \Gamma e^{i \eta_{31}} & \kappa' e^{i \eta_{32}} & \tau_2
    \end{pmatrix}.
    \label{general3x3}
\end{equation}

The elements of $U_{3\times 3}$ are expressed in polar form and the $\eta_{ij}$ are, for the moment, assumed to be arbitrary. After this, we can eliminate the variables $\kappa'$ and $\Gamma$ by enforcing unitarity. Furthermore, this unitarity condition requires that the columns be orthogonal. This imposes a condition on the phases. Since the elements are expressed in polar coordinates and the columns must be orthogonal, the simplest condition the phases can satisfy under the additional assumption of symmetry is $\eta_{12} + \eta_{23} + \eta_{31} = \pi$. This introduces a relative  overall minus sign to the columns so that orthogonality condition can be enforced. Finally, we assume that the transmission coefficients are symmetric in the rings (reducing our system to a single $\tau$). Under these assumptions, $U_{3 \times 3}$, which represents the central 3-mode junction (as seen in the lower right of \Fig{fig:rr_circuit}), reduces to:

\begin{equation}
        U_{3 \times 3} = 
        \begin{pmatrix}
    \tau_1 & -\sqrt{2\tau_1(1-\tau_1)} & \tau_1-1\\
    -\sqrt{2\tau_1(1-\tau_1)} & 1-2\tau_1 & \sqrt{2\tau_1(1-\tau_1)}\\
    \tau_1-1 & \sqrt{2\tau_1(1-\tau_1)} & \tau_1
    \end{pmatrix}
    \label{2taumatrix}
\end{equation}

We use $U_{3 \times 3}$ to represent the central junction of our system and $U_{2 \times 2}$ to represent both outer junctions and enforce the matching boundary conditions to yield the overall $S$ matrix whose components are complicated algebraic functions of the remaining tunable parameters. By the choices we make here, these remaining tunable parameters are the round-trip phase shifts of the rings $\{\theta_j\}$ as well as the transmission coefficients $\{\tau_j\}$ at each junction. We therefore suggestively write

\begin{equation}
    S_{ij} = S_{ij}(\vec{\tau}, \vec{\theta})
\end{equation}

\noindent to indicate this fact. The literature suggests that there are fabrication techniques which can realize this $3 \times 3$ tunable device \cite{Skryabin:2020}. This $S$ matrix can then be used to relate the input and output states. We proceed with this in the next section.

{\flushleft{\it Results:}}

To discuss the results of this device, we will define the probability of successfully detecting the desired output state on the $b$ output mode as the heralding success probability $P_{click}$. Furthermore, we will define the fidelity of the output state compared to the desired NOON state output across the $a$ and $c$ output modes upon a successful heralding as $F_{NOON}$. For the results in this paper we assume the device to be lossless and we work in the continuous wave (cw) limit. That is to say, the optical path length of the device is much less than the coherence length of the light. These MRR-based devices are extremely robust around $\theta=\pi$. We have investigated the stability, loss, and backscattering of the Hong-Ou-Mandel effect in MRR-based devices extensively and refer the reader to our previous work for additional information \cite{Kaulfuss:2023Identical,Alsing_Hach:2017a,Kaulfuss:2023Backscatter,eHOM_Alsing:2022}. We also assume the generation of Fock state inputs for this device.

Beginning with a single photon Fock state input on each waveguide and heralding a success on the vacuum output state on the central mode, $\ket{0}_b$, yields: 

\begin{widetext}
    \begin{multline}
        |\psi_{in}\rangle = \ket{1,1,1}_{a,b,c} \xrightarrow{S} |\psi_{out}\rangle = \ket{0}_b\otimes \bigg[C_{0,0,3}(\Vec{\tau}, \vec{\theta})\ket{0}_a\ket{3}_c + C_{1,0,2}(\Vec{\tau}, \vec{\theta})\ket{1}_a\ket{2}_c +\\
         C_{2,0,1}(\Vec{\tau}, \vec{\theta})\ket{2}_a\ket{1}_c + C_{3,0,0}(\Vec{\tau}, \vec{\theta})\ket{3}_a\ket{0}_c \bigg].
    \end{multline}
    \label{3NOONherald0eq}
\end{widetext}

\noindent Where $\vec{\tau}$ and $\vec{\theta}$ are to denote the collection of tunable independent parameters consistent with our derivation of the $S$ matrix above. Through optimization of the tuning parameters, we are able to minimize the amplitudes of the $\ket{1,0,2}_{a,b,c}$ and $\ket{2,0,1}_{a,b,c}$ output states to numerically zero. This means that for an entire manifold of the $\tau_j$ and $\theta_j$ combinations when $\ket{0}_b$ is measured there is certainly an output state of the form $C_{0,0,3}|0,0,3\rangle + C_{3,0,0}|3,0,0\rangle$; we can then use state fidelity to enforce $F_{NOON} = 1$, so that, in particular, $C_{3,0,0} = C_{0,0,3} = 1/\sqrt{2}$. The probability of heralding the vacuum on the $b$ output mode is $P_{click}=\frac{4}{9}$. This is also the maximum probability predicted for a 3-photon NOON state with unitary matrix optimization \cite{DmitriNOON}. This is similar to other schemes discussed in the introduction that rely on the detection of zero photons in an ancillary mode. However, it is difficult experimentally to herald on a vacuum state output. Therefore, to make our NOON state generation scheme more experimentally viable, we now explore heralding on single photon detection in the $b$ mode, $\ket{1}_b$, and input an additional photon on the central waveguide:

\begin{widetext}
    \begin{multline}
        |\psi_{in}\rangle = \ket{1,2,1}_{a,b,c} \xrightarrow{S} |\psi_{out}\rangle = \ket{1}_b\otimes \bigg[C_{0,1,3}(\Vec{\tau}, \vec{\theta})\ket{0}_a\ket{3}_c + C_{1,1,2}(\Vec{\tau}, \vec{\theta})\ket{1}_a\ket{2}_c+ \\
         C_{2,1,1}(\Vec{\tau}, \vec{\theta})\ket{2}_a\ket{1}_c + C_{3,1,0}(\Vec{\tau}, \vec{\theta})\ket{3}_a\ket{0}_c \bigg]
    \end{multline}
    \label{3NOONherald1eq}
\end{widetext}

We assume access to number resolving detectors yielding the ability to herald on a single photon output in the $b$ mode. Number resolving detectors have been shown for up to 100 photons with SNSPD band-edge detectors \cite{Eaton:2023,Risheng:2023}. In this single photon heralding scheme we are also able to completely minimize the non-NOON output states to yield $F_{NOON}=1$ with $P_{click}=\frac{8}{27}$, or a probability of 29.6\%. This means that nearly 30\% of the time with this $\ket{1,2,1}_{a,b,c}$ input state and certain $\tau$ and $\theta$ combinations, a single photon will be detected at the $b$ output mode and when this single photon detection occurs there is a 3-photon NOON state across the $a$ and $c$ output modes with absolute certainty whose location can be tracked by time-of-flight methods. 

\begin{figure}[H]
	\centering
	\includegraphics[width=3in,keepaspectratio]{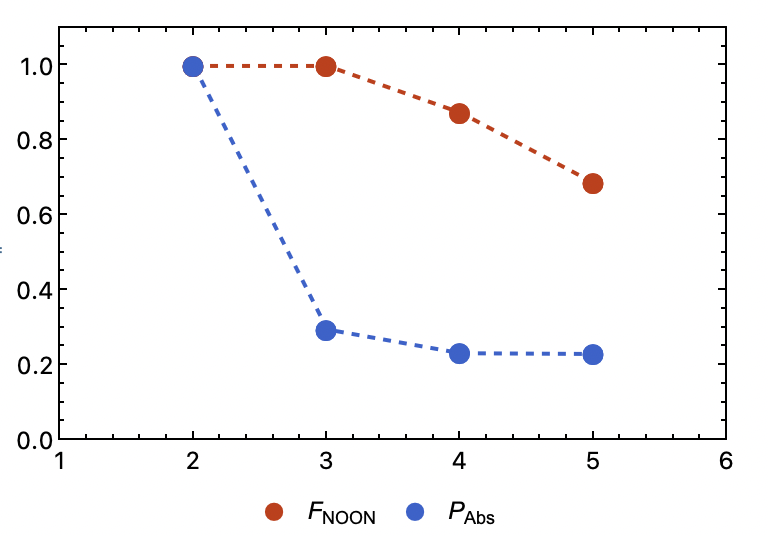}
	\caption{We show the values for $P_{Click}$ and $F_{NOON}$ for High-NOON state $N$ values 2-5. The 2 photon NOON state can be obtained deterministically using a 50/50 beam splitter. For N values 3, 4, and 5 we assume our MRR-based device upon heralding a single photon on the central output mode $b$. The parameter values for $N = 3$ are $\tau_0 \approx 0.52$, $\tau_1 \approx 0.54$, and $\theta \approx \pi$, for $N = 4$ are $\tau_0 \approx 0.50$, $\tau_1 \approx 0.56$, and $\theta \approx \pi$, and for $N = 5$ are $\tau_0 \approx 0.25$, $\tau_1 \approx 0.66$, and $\theta \approx \pi$}
    \label{fig:allProbs}
\end{figure}

This device can be easily expanded to High-NOON states by increasing the number of photons input into the $b$ mode of the device, while maintaining single photon Fock states on the $a$ and $c$ inputs and continuing to herald on a single photon output in the $b$ mode. There have been multiple experimental demonstrations of the generation of higher photon number Fock states \cite{Cooper:2013,Rivera:2023,Hua:2020,Pryde:2003}. For example, by inputting $\ket{1,3,1}_{a,b,c}$ and continuing to herald on a single photon output in the $b$ mode, a 4-photon NOON state can be achieved with $P_{click}=0.23$ and $F_{NOON}=0.85$. This trend continues for higher $N$ High-NOON states, with  both $P_{click}$ and $F_{NOON}$ decreasing as $N$ increases because there are increasing numbers of `accidental' states that cannot be completely eliminated through minimization with a consistent set of tunable device parameters. Note that relaxing the constraints of the system to increase the number of tunable parameters may allow unit fidelity to be maintained for higher $N$ High-NOON states, which may be the subject of future work. Values for $P_{Click}$ and $F_{NOON}$ and for High-NOON states up to N=5 are are shown in Fig. \ref{fig:allProbs}. 

Additionally, as in, e.g., \cite{Alsing_Hach:2019}, one can easily show that the parameter space has dimension $d > 0$ (i.e., consists of a topology finer than a mere set of discrete points) according to the same general scheme. This suggests, as with objects in the class of circuits generated by the general ring resonator scheme, that the circuit element is characterized by a set of operating parameter values which enables robust tuning and handling of errors. That is to say there exists an entire manifold of solutions for $\tau$ and $\theta$ combinations that yield the optimal NOON state output with maximum probability. 


We have designed a device that allows for the non-destructive probabilistic heralding of High-NOON states. NOON states are vitally important for quantum sensing applications, allowing for a quantum improvement in sensitivity, leading to more precise measurements \cite{LeeKokDowling:2002}.

With Fock state generation and single photon detection capabilities on the ancilla mode of the 2 MRR device and the correct choice of tunable device parameters, one can achieve 100\% certainty of a 3-photon NOON state upon a successful heralding single photon detection on the middle mode, which occurs with a probability of 29.6\%. A 3-photon NOON state output is guaranteed upon a successful heralding event without destroying the NOON state output and the NOON state can be tracked using time-of-flight methods. The device and scheme reported can be generalized for High-NOON states, by inputting higher $N$ Fock states into the central waveguide, with decreasing fidelity due to the increased number of `accidental' states that cannot be eliminated by optimization of tunable parameters. This MRR based device and heralding scheme allows for the generation of High-NOON states that can be tracked non-destructively via heralding and time of flight calculations with high probability.

\begin{acknowledgments} 
\label{Acknowledgements}
R.S. and E.E.H. acknowledge support from the United States Air Force Research Laboratory (AFRL) Summer Faculty Fellowship Program for providing support for this work. PLK acknowledges support from the AFRL through the QUEST contract (FA8750-23-C-0053).  
The US Government is authorized to reproduce and distribute reprints for Government Purposes notwithstanding any copyright notation thereon. The views and conclusions contained herein are those of the authors and do not necessarily represent the official policies or endorsements, either explicit or implied of the United States Air Force, the Air Force Research Laboratory or the U.S. Government. 
\end{acknowledgments}

\bibliography{references}

\end{document}